\documentstyle[twocolumn]{jpsj}
\input{epsf.tex}
\newcommand{\bm}[1]{\mbox{\boldmath$#1$}}
\title{Spin-Gap Phase in the One-Dimensional $t$-$J$-$J'$ Model}
\author{Masaaki \sc{Nakamura}
\footnote{E-mail address: masaaki@stat.phys.kyushu-u.ac.jp}}
\inst{Department of Physics, Kyushu University, Fukuoka 812-81\\
Department of Physics, Tokyo Institute of Technology,
Oh-Okayama, Meguro-ku, Tokyo 152}
\kword{Tomonaga-Luttinger liquid, spin gap, backward scattering,
renormalization group, singlet-triplet level-crossing,
twisted boundary conditions,
factorized wave function, two-electron problem}
\def\runauthor{Masaaki \sc{Nakamura}}
\def\runtitle{Spin-gap phase in the one-dimensional $t$-$J$-$J'$ model}
\recdate{\today}
\abst{
 The spin-gap phase of the one-dimensional $t$-$J$-$J'$ model
 is studied by the level-crossing of
 the singlet and the triplet excitation spectra.
 The phase boundary obtained between the Tomonaga-Luttinger
 and the spin-gap phases is remarkably consistent with the analytical results
 at the $J,J'\rightarrow 0$ and the low-density limits
 discussed by Ogata {\it et al}.
 The spin-gap phase has a single domain in the phase diagram
 even if the spin gap opens at half-filling.
 The phase boundary coincides with
 the $K_{\rho}=1$ line where the Tomonaga-Luttinger liquid
 behaves as free electrons, in the low-density region.
 The relation between our method and the solution of the
 two-electron problem is also discussed.}
\begin{document}
\sloppy
\maketitle 
The discovery of quasi one-dimensional (1D) cuprates
has stimulated the study of 1D strongly correlated electron systems.
The existence of a spin gap is an important factor
to be considered in the investigation of superconductivity.
The effect of frustration was
considered to be a candidate causing a spin gap
in the realistic parameter region.
However, it was very difficult to determine the phase boundary between
the Tomonaga-Luttinger (TL) and the spin-gap phases,
due to the singular behavior of the spin gap near the critical point.
Recently, the author with Nomura and Kitazawa have developed
a new method to treat the spin gap,
that is, the singlet-triplet level-crossing method
with the twisted boundary conditions.\cite{Nakamura-N-K}
In this letter, we will apply this method to the 1D $t$-$J$-$J'$ model
and show that the result is consistent with the exact results
in the low-density and the $J,J'\rightarrow 0$ limits,
and we will clarify the spin-gap phase of this model.

The Hamiltonian of the 1D $t$-$J$-$J'$ model is written
in the subspace of no doubly occupied sites as
\begin{eqnarray}
{\cal H}&=&
 -t\sum_{i\sigma}(c^{\dag}_{i\sigma} c_{i+1\sigma}+\mbox{H.c.})\nonumber\\
 &&+\sum_{l=1,2}\sum_{i}J^{(l)}(\bm{S}_i\cdot\bm{S}_{i+l}-n_i n_{i+l}/4),
  \label{eqn:t-J-J'}
\end{eqnarray} 
where $J^{(1)}=J, J^{(2)}=J'$.
We also introduce a parameter $\alpha$ for the
strength of the frustration given by $\alpha\equiv J'/J$.
At half-filling ($n=1$),
this model becomes an $S=1/2$ frustrated spin chain.
In this case, the ground state
at $\alpha=1/2$ is the two-fold degenerate dimer state with a spin gap,
and the ground state energy density is $-3/4J$.
\cite{Majumder-G,Shastry-S,Broek}
Okamoto and Nomura have argued, using the level crossing method,
that the fluid-dimer transition occurs at $\alpha_{c}=0.2411$.\cite{Okamoto-N}
Upon doping of holes, the system may become metallic,
and the spin gap is reduced\cite{Sano-T} but persists for the finite doping.
The phase diagram of this model for $n\neq 1$ at $\alpha=1/2$,
using the exact diagonalization,
was obtained by Ogata, Luchini and Rice.\cite{Ogata-L-R}
They also made an important achievement regarding
the critical points for some limits.

We briefly review their method to obtain
the critical points in the following two limits:
in the limit of $J,J'\rightarrow 0$,
the spin part of this model can be mapped onto the case of $n=1$,
using the factorized wave function\cite{Ogata-L-R,Ogata-S},
\begin{subequations}
\begin{eqnarray}
 J_{\rm eff}&=&J\langle n_i n_{i+1}\rangle_{\rm SF}+
 J'\langle n_i (1-n_{i+1})n_{i+2}\rangle_{\rm SF},\\
 J'_{\rm eff}&=&J'\langle n_i n_{i+1}n_{i+2}\rangle_{\rm SF},
\end{eqnarray}
\end{subequations}
where $\langle\cdots\rangle_{\rm SF}$ indicates the expectation value of the
non-interacting spinless fermion.
The effective ratio of the frustration $\alpha_{\rm eff}$
is then obtained as
\begin{equation}
 \alpha_{\rm eff}(n,\alpha)=\left[
   \frac{(1+1/\alpha)n^{\,2}-s_2^{\,2}-s_1^{\,2}/\alpha}
   {n^3-(2s_1^{\,2}+s_2^{\,2})n+2s_1^{\,2}s_2}-1\right]^{-1},
  \label{eqn:eff_alpha}
\end{equation}
where $s_l\equiv \sin(l\pi n)/l\pi n$.
We can obtain the critical density $n_c$ where the spin gap vanishes,
by comparing eq.(\ref{eqn:eff_alpha}) with the value $\alpha_c=0.2411$
obtained by Okamoto and Nomura.\cite{Okamoto-N}
For $\alpha=1/2$, $n_c=0.7433$.
On the other hand, for the low density limit,
the critical value of $J/t$ where the singlet pair forms a bound state,
can be analytically obtained by solving the two-electron problem.
The solution can be derived
by adding a Hubbard interaction
${\cal H}_U\equiv U\sum_i n_{i\uparrow}n_{i\downarrow}$
to the Hamiltonian, relaxing the constraint,
and setting $U=\infty$ at the end of the calculation\cite{Lin}.
The obtained result is
\begin{equation}
 \frac{J_c}{t}=\frac{1+2\alpha-\sqrt{1+4\alpha^2}}{\alpha}.
 \label{eqn:two_elec}
\end{equation}
It is not trivial that this critical value is equivalent to
the spin-gap phase boundary. We will discuss this point later.

The critical point in the remaining region
can be determined in the following way.\cite{Nakamura-N-K}
In this letter we argue within the scheme of the bosonization theory.
The low energy behavior of 1D electron systems is described by
the sine-Gordon model\cite{Solyom,Schulz}
\begin{equation}
 {\cal H}={\cal H}_{\rho} + {\cal H}_{\sigma}
  + \frac{2 g_{1\perp}}{(2\pi a)^2}\int dx \cos(\sqrt{8}\phi_{\sigma})
  \label{eqn:effHam}.
\end{equation}
Here $a$ is a short-distance cutoff, $g_{1\perp}$ is the backward
scattering amplitude and for $\nu=\rho,\sigma$
\begin{equation}
  {\cal H}_{\nu}=\frac{1}{2\pi}\int dx
  \left[
   v_{\nu}K_{\nu}\left(\frac{\partial \theta_{\nu}}{\partial x}\right)^2
   +\frac{v_{\nu}}{K_{\nu}}
   \left(\frac{\partial \phi_{\nu}}{\partial x}\right)^2
  \right]\label{eqn:Gaussian},
\end{equation}
where $v_{\nu}$ and $K_{\nu}$ are
the velocity and the Gaussian coupling, respectively,
for the charge ($\nu=\rho$) and the spin ($\nu=\sigma$) sectors.
In the TL phase ($g_{1\perp}>0$), the parameters $K_{\sigma}$ and $g_{1\perp}$
are renormalized as
$K^{*}_{\sigma}=1$ and $g_{1\perp}^{*}=0$, reflecting the SU(2) symmetry.
The phase fields are defined as\cite{Schulz,Haldane}
\begin{eqnarray}
 \lefteqn{\phi_{\nu}(x),\theta_{\nu}(x)=}\label{eqn:phase_fields}\\
&&  \mp\frac{i\pi}{L}\sum_{p\neq 0}\frac{1}{p}e^{-i a|p|/2-ipx}
  \left[\nu_R(p)\pm \nu_L(p)\right]
  \mp n_{\nu},m_{\nu}\frac{\sqrt{2}\pi x}{L},\nonumber
\end{eqnarray}
where $\nu_{r}$ is the charge ($\nu=\rho$) or the spin ($\nu=\sigma$)
density operator, which satisfy the boundary conditions,
\begin{subequations}
\begin{eqnarray}
 \phi_{\nu}(x+L)&=&\phi_{\nu}(x)-\sqrt{2}\pi n_{\nu},\\
 \theta_{\nu}(x+L)&=&\theta_{\nu}(x)+\sqrt{2}\pi m_{\nu}.
\end{eqnarray}\label{eqn:BC_phase}
\end{subequations}
These phase fields have the relation
$\left[\phi_{\nu}(x),\partial_{x'}\theta_{\nu}(x')\right]=i\pi\delta(x-x')$.
The quantum numbers are defined by
the total number operators (measured with respect to the ground state)
$N_{r\sigma}$ for right and left going particles ($r=R,L$) of spin $\sigma$
\cite{Schulz,corresp}
\begin{subequations}
\begin{eqnarray}
 n_{\nu}&=&
  [(N_{R\uparrow}+N_{L\uparrow})\pm (N_{R\downarrow}+N_{L\downarrow})]/2,\\
 m_{\nu}&=&
  [(N_{R\uparrow}-N_{L\uparrow})\pm (N_{R\downarrow}-N_{L\downarrow})]/2.
\end{eqnarray}\label{eqn:q_num}
\end{subequations}
Here the upper and lower sign refer to charge and spin degrees of freedoms,
respectively.
The selection rule among them for $N=4l+2$ ($l$: integer) electrons
in periodic boundary conditions is\cite{Haldane,Woynarovich}
\begin{equation}
 (-1)^{m_{\rho}\pm m_{\sigma}}=(-1)^{n_{\rho}\pm n_{\sigma}}.
  \label{eqn:sel_rule}
\end{equation}
The finite-size corrections for the excitation energy and momentum
of the system with length $L$ are described by
\begin{eqnarray}
 E-E_0&=&
  \frac{2\pi v_{\rho}}{L}x_{\rho}+\frac{2\pi v_{\sigma}}{L}x_{\sigma},\\
 P-P_0&=&\frac{2\pi}{L}(s_{\rho}+s_{\sigma})+2m_{\rho}k_{F},
\end{eqnarray}
where $k_F=\pi n/2$ is the Fermi wave number,
$x_{\nu}\equiv(m_{\nu}^{\ 2}K_{\nu}+n_{\nu}^{\ 2}/K_{\nu})/2$
and $s_{\nu}\equiv m_{\nu}n_{\nu}$ are the scaling dimension and
the conformal spin, respectively, for each sector.
The corresponding operator is given by
$\exp[i\sqrt{2}(m_{\nu}\phi_{\nu}+n_{\nu}\theta_{\nu})]$.
The boson representation of the fermion operator is
\begin{equation}
 \psi_{r,\sigma}=\frac{1}{\sqrt{2\pi a}}
  e^{i r k_F x}e^{i/\sqrt{2}\cdot
 [r(\phi_{\rho}+\sigma \phi_{\sigma})-\theta_{\rho}-\sigma \theta_{\sigma}]},
  \label{eqn:fermion_op}
\end{equation}
where $r=R,L$ and $\sigma=\uparrow,\downarrow$ refer to $+,-$ in that order.

Next we extract the singlet and the triplet excitation spectra
$[m_{\rho}=n_{\rho}=0,(m_{\sigma},n_{\sigma})=(1,0),(0,\pm 1)]$.
The corresponding operators are
$\sqrt{2}\cos\sqrt{2}\phi_{\sigma}$ for the singlet and
$\sqrt{2}\sin\sqrt{2}\phi_{\sigma}, 
\exp(\mp i\sqrt{2}\theta_{\sigma})$ for the triplet state.\cite{Giamarchi-S}
Using eqs. (\ref{eqn:BC_phase}) and (\ref{eqn:fermion_op}),
it can be seen that the fermion operator should satisfy
the anti-periodic boundary conditions (APBC) for these excitations:
$\psi_{r,\sigma}(x+L)=-\psi_{r,\sigma}(x)$.
This indicates that APBC plays a role
in the elimination of the induced charge excitations.\cite{twisting}

The effect of the backward scattering term with the coupling $g_{1\perp}$ is
considered using a renormalization group analysis
under the change of the cutoff $a\rightarrow e^{dl}a$,
then the scaling dimensions of the singlet ($x_{\sigma s}$)
and the triplet ($x_{\sigma t}$) excitations split logarithmically
by the marginally irrelevant coupling
$y_0(l)\equiv g_{1\perp}/\pi v_{\sigma}$ as
\cite{Giamarchi-S,Affleck-G-S-Z}
\begin{subequations}
\begin{eqnarray}
x_{\sigma s}&=&\frac{1}{2}+\frac{3}{4}\frac{y_0}{y_0\ln L+1},\\
x_{\sigma t}&=&\frac{1}{2}-\frac{1}{4}\frac{y_0}{y_0\ln L+1},
\end{eqnarray}\label{eqn:sclngdim}
\end{subequations}
where $y_0$ is the bare coupling $y_0\equiv y_0(l=0)$.
In the TL region where the backward scattering is repulsive ($y_0>0$),
$g_{1\perp}$ is renormalized to $0$.
On the other hand, when the backward scattering becomes attractive ($y_0<0$),
$g_{1\perp}$ is renormalized to $-\infty$ and a spin gap appears.
At the critical point $y_0=0$,
there are no logarithmic corrections.
Thus, the critical point is determined with high-accuracy
by the intersection of these spectra.
This way of determination of the spin-gap phase boundary is equivalent to
the spin-gap generation in the theory of TL liquids.\cite{Solyom}

\begin{figure}
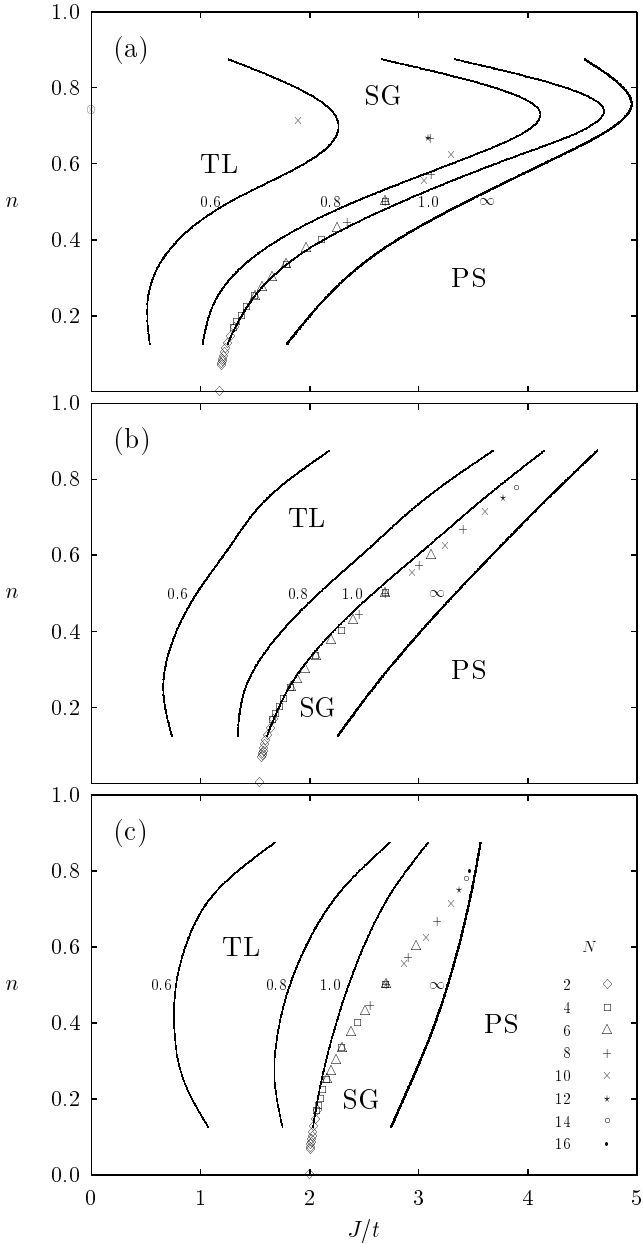

 \epsfxsize=2.8in \leavevmode \epsfbox{fig1a.epsi}\\
 \epsfxsize=2.8in \leavevmode \epsfbox{fig1b.epsi}\\
 \epsfxsize=2.8in \leavevmode \epsfbox{fig1c.epsi}
\caption{Phase diagram of the 1D $t$-$J$-$J'$ model at
 (a) $\alpha=1/2$, (b) $\alpha=0.2411$, (c) $\alpha=0$
  (TL: TL liquid, SG: spin-gap phase, PS: phase-separated state).
 The spin-gap phase boundary flows into the critical point obtained
 by the analysis of Ogata {\it et al.}\cite{Ogata-L-R}
 (plotted by $\otimes$ in (a)).
 The contour lines of $K_{\rho}$ are determined
 by the data of the $L=16$ system,
 based on the analysis of Ogata {\it et al.}\cite{Ogata-L-S-A,Ogata-L-R}}
 \label{fig:phsdgrm}
\end{figure}

Fig. 1(a) shows the phase diagram of eq.(\ref{eqn:t-J-J'}) at $\alpha=1/2$.
The exponent $K_{\rho}$ is calculated in the $L=16$ system
in a similar manner as shown by Ogata {\it et al}.,
which was used for the $t$-$J$(-$J'$) model.
\cite{Ogata-L-R,Ogata-L-S-A,Nakamura-N}
The phase boundary starts from the critical value of the low-density limit,
and bends at $n\sim2/3$.
It then flows into the critical point of the $J,J'\rightarrow 0$ limit.
Thus the spin-gap phase appears near half-filling,
and in the large $J/t$ region as in the $t$-$J$ model,
but it has a single domain in the phase diagram.
Ogata {\it et al.}
estimated that the $J/t$ dependence of the spin-gap phase
in the low-doping region is small,
and observed the spin gap in the large $J/t$ region
at $n=2/3$ using a finite-size scaling method.\cite{Ogata-L-R}
These estimations are almost consistent with our results.

Another interesting feature of the phase diagram is that
the spin-gap phase boundary overlaps
the $K_{\rho}=1$ line where the TL liquid behaves as free electrons
in the low density region.
This point will be clarified later
in connection with the two-electron problem.

In the theory of TL liquids,
both the singlet and the triplet superconducting correlations (SS, TS)
have the same critical exponent $1/K_{\rho}+1$,\cite{Solyom}
but TS is dominant if the logarithmic corrections
are taken into account.\cite{Giamarchi-S}
However, in the presence of the spin gap,
SS is enhanced to $1/K_{\rho}$ and TS decays exponentially.\cite{Ogata-L-R}
In the case of the $t$-$J$ model ($\alpha=0$),
the spin-gap phase boundary lies in the $K_{\rho}>1$ region
\cite{Nakamura-N-K,Ogata-L-S-A}(see Fig. \ref{fig:phsdgrm}(c)).
On the other hand, at $\alpha=1/2$ it lies in the $K_{\rho}<1$ region,
so that there is no TS region in the latter case.

The phase diagrams for the intermediate values of $\alpha$
can be constructed in the same way.
As an example, we show the phase diagram
at $\alpha=\alpha_c$ in Fig. \ref{fig:phsdgrm}(b).
As $\alpha$ is increased from $0$,
the spin-gap phase boundary and
the phase-separation boundary \cite{lines} shift
towards the small $J/t$ side in the low-density region.
On the other hand, in the high-density region,
the two lines shift to the larger $J/t$ side.
However, after $\alpha$ exceeds the value $\alpha_c$,
the spin-gap phase boundary bends and flows into the point $(J/t,n)=(0,n_c)$.
The phase separation boundary also goes to the $J/t\rightarrow 0$ side
following the behavior of the spin-gap phase boundary.
\begin{figure}
 \epsfxsize=2.8in \leavevmode \epsfbox{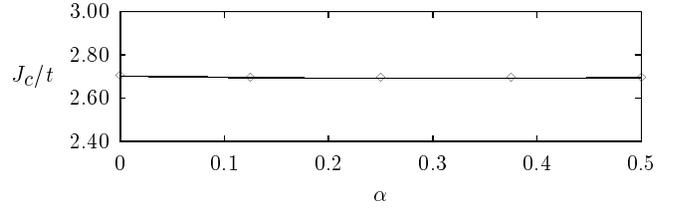}
\caption{Critical point $J_c/t$ versus
 the strength of the frustration $\alpha$
 in the $L=16$ system at $n=1/2$.
 The critical value is almost constant ($J_c/t\sim2.7$).
 This behavior can be explained by applying the {\it g-ology}.}
 \label{fig:qf}
\end{figure}
This behavior indicates that the phase-separated region is
always situated on the border of the spin-gap phase.

In spite of the deformation of the phase diagram,
the critical value $J_c/t$ at the quarter-filling ($n=1/2$)
is almost independent of the strength of the frustration $\alpha$,
and is kept at $J_c/t\sim 2.7$ as shown in Fig. \ref{fig:qf}.
Let us consider the reason for this
using an argument based on the {\it g-ology}.\cite{Solyom}
In order to apply the {\it g-ology},
we add the on-site Coulomb term
${\cal H}_U$ to eq.(\ref{eqn:t-J-J'}) and relax the constraint.
The original Hamiltonian is restored when we set $U=\infty$.
The interaction term in eq.(\ref{eqn:t-J-J'}) is divided into
the $XY$ and Ising terms as
\begin{subequations}
\begin{eqnarray}
 {\cal H}^{(l)}_{XY}&=&
  \frac{J^{(l)}}{2}\sum_i(S^+_iS^-_{i+l}+S^-_iS^+_{i+l}),\\
 {\cal H}^{(l)}_{\rm Ising}
  &=&-\frac{J^{(l)}}{2}\sum_i
  (n_{i\uparrow}n_{i+l\downarrow}+n_{i\downarrow}n_{i+l\uparrow}).
\end{eqnarray}
 \label{eqn:J_terms}
\end{subequations}
Since the {\it g-ology} is appropriate for the weak coupling case,
we consider the $l=2$ terms of eq.(\ref{eqn:J_terms})
as corrections to the $t$-$J$ model
which belongs to the universality class of the TL model.
Then their contributions to the $g$-parameters,
which are related to the spin-gap generation, are identified as
\cite{Solyom}
\begin{equation}
 \delta g_{1\perp}=\delta g_{1\parallel}
  =-J'(1+\cos 4 k_F).
  \label{eqn:correction}
\end{equation}
Note that the $g$-parameters are redefined as
$g_{1\parallel}-g_{2\parallel}+g_{2\perp}\rightarrow g_{1\parallel}$.
For the quarter-filling, eq.(\ref{eqn:correction}) vanishes,
so that the $l=2$ terms of eq.(\ref{eqn:J_terms})
do not affect the renormalization flow of the spin part.
Thus the frustration does not change the critical point
at the quarter-filling within the scheme of the {\it g-ology}.

Finally, we discuss the equivalence between the level-crossing method
and the solution of the two-electron problem,\cite{Lin}
and the accordance between the spin-gap phase boundary
and the $K_{\rho}=1$ line in the low-density region.
In the low-density limit,
the many-body problem can be reduced to a two-body problem.
The wave function of the two-electron system is given by
\begin{equation}
 \Psi=\sum_{ij}\Phi(i,j)c_{i\uparrow}^{\dag}c_{j\downarrow}^{\dag}|vac\rangle,
\end{equation}
where $\Phi(i,j)=\Phi(j,i)$ for the singlet
and $\Phi(i,j)=-\Phi(j,i)$ for the triplet state.
It is well-known that for a two-body problem the ground state is
a singlet state as far as the bottom of the energy band has no degeneracy.\cite{Slater-S-K,Yoshida}
\begin{figure}
 \epsfxsize=2.8in \leavevmode \epsfbox{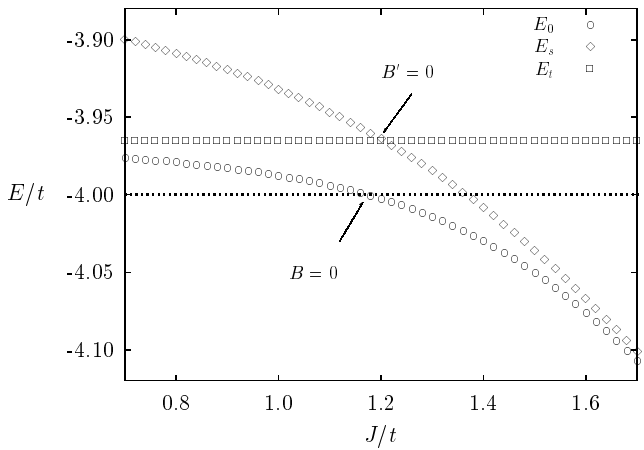}
\caption{Ground state energy ($E_0$)
 and the first singlet and the triplet excitation
 spectra ($E_s, E_t$) of the two-electron system
 with length $L=24$ at $\alpha=1/2$.
 $E_s$ and $E_t$ are calculated using anti-periodic boundary conditions.
 The marked intersections will coincide in the $L\rightarrow\infty$ limit.}
 \label{fig:two_ele}
\end{figure}
For periodic boundary conditions,
the singlet ground state energy is given by $E_0=-4t+B$
where $B$ is the bound energy,
and the critical value where $B=0$ is given by eq.(\ref{eqn:two_elec})
without size dependence.
However, for APBC, the bottom of the energy band is degenerate,
so that the triplet state can be the ground state.
One can show that there is no effect of $J^{(l)}$
for the triplet state.
Then the lowest energy of the triplet state for APBC
is always $E_t=-4t+\delta(L)$
where $\delta(L)\equiv 8t \sin^2 (\pi/2L)$.
On the other hand,
the energy for the singlet state for APBC is given by $E_s=-4t+\delta(L)+B'$
where $B'$ is the bound energy.
Then the level-crossing between the two levels takes place at $B'=0$
(see Fig. \ref{fig:two_ele}).
In the thermodynamic limit, $\delta(L\rightarrow\infty)=0$
and the difference between boundary conditions should vanish.
This means that the critical points for $B=0$ and $B'=0$
will be the same in the thermodynamic limit.
In this way, it turns out that the singlet-triplet level-cross point
is nothing but the solution of the two-electron problem ($B=0$)
in the low-density limit.

In the TL liquid theory
electrons behave as free particles at $K_{\rho}=1$,
so that the spin-gap phase boundary,
where the bound energy becomes $0$,
should overlap the $K_{\rho}=1$ line in the low-density limit.
Moreover, since the system in the $J<J_c$ region is equivalent to
the non-interacting spinless fermion system ($K_{\rho}=0.5$),
the contour lines for $0.5<K_{\rho}<1$ will focus on the point $(J_c/t,0)$.

In conclusion, we studied the spin-gap phase of the 1D $t$-$J$-$J'$ model
using the singlet-triplet level-crossing method
with the twisted boundary conditions.
The obtained results show remarkable consistency
with the exactly known results.
The spin-gap phase appears near the half-filling as expected,
and also appears as a precursor of the phase separation
as in the $t$-$J$ model.
The spin-gap phase has a single domain in the phase diagram.
The overlap between the spin-gap phase boundary and the $K_{\rho}=1$ line
in the low density region is explained in the connection with
the two-electron problem.

The author is very grateful to A. Kitazawa and K. Nomura
for helpful discussions and critical reading of the manuscript.
The author also thanks M. Ogata, K. Okamoto and H. Shiba for valuable comments.
This work is partially supported by a Grant-in-Aid for
Scientific Research (C) No. 09740308 from the Ministry of Education,
Science, Sports and Culture, Japan.
The computation work was done
using the facilities of the Supercomputer Center,
Institute for Solid State Physics, University of Tokyo.


\begin{thebibliography}{99}
\bibitem{corresp}
 The quantum numbers of Woynarovich's notation\cite{Woynarovich}
 can be read as
  $\Delta N_c =
  N_{R\uparrow}+N_{L\uparrow}+N_{R\downarrow}+N_{L\downarrow},
 \Delta N_s = N_{R\downarrow}+N_{L\downarrow},
  2D_c        =  N_{R\uparrow}-N_{L\uparrow},
  2D_c + 2D_s =  N_{R\downarrow}-N_{L\downarrow}$,
 and parameters for the Gaussian model are given by
 $m_{\rho}=2D_c+D_s, n_{\rho}=\Delta N_c/2,
 m_{\sigma}=-D_s, n_{\sigma}=\Delta N_c/2-\Delta N_s$.
\bibitem{Woynarovich}
  F. Woynarovich: J. Phys. A {\bf 22} (1989) 4243.
 In this paper, the selection rule is derived
 from the Bethe-ansatz result of the $U\rightarrow\infty$
 limit of the Hubbard model.
\bibitem{twisting}
 In fact, we can determine the critical point without twisting
 the boundary conditions,
 but the excitation energies become higher and
 the finite-size effect becomes larger.
 The elimination of the charge excitation is also important
 in the identification of the universality class of the transition.
\bibitem{lines}
In contrast to the spin-gap phase boundary,
the finite-size effect of the phase-separation boundary is large,
and its systematic evaluation is difficult,
so that the remaining ambiguity should be considered.
\end{thebibliography}
\end{document}